# STRUCTURE OF THE UNILAMELLAR DIMYRISTOYLPHOSPHATIDYLCHOLINE VESICLE. A SMALL-ANGLE NEUTRON SCATTERING STUDY.

E.V.Zemlyanaya[1], M.A.Kiselev[2], V.K. Aswal[3]

[1]*Laboratory of Information Technologies, JINR, Dubna 141980, Russia*

[2]*Laboratory of Neutron Physics, JINR, Dubna 141980, Russia*

[3]*Paul Scherrer Institute, CH-5232, Villigen, Switzerland*

Running Tittle: Structure of unilamellar vesicles

Corresponding author: E.V.Zemlyanaya

*Laboratory of Information Technologies, JINR, Dubna 141980, Russia*

Tel.: 007 (09621) 647-28

E-mail: elena@jinr.ru

**Abstract**

On the basis of the separated form-factors, a code for fitting the small-angle neutron scattering spectra of the polydispersed vesicle population have been developed. Vesicle and membrane bilayer parameters are analyzed for various hierarchical models of the neutron scattering length density across the membrane. It is shown that hydration of vesicle can be described by a linear distribution function of water molecules. For the first time, the average radius and polydispersity of the vesicle population, thickness of the membrane bilayer, thickness of hydrophobic and hydrophilic parts of bilayer, and water distribution function have been calculated from the small-angle experiment, without additional methods, The results obtained at two different spectrometers, are discussed. The appropriate conditions of the SANS experiment on vesicles are formulated as a necessity to collect the SANS curve in the region of scattering vectors from $q_{min}=0.0033$Å$^{-1}$ to $q_{max}=0.56$Å$^{-1}$.

1. **INTRODUCTION**

Research in the structure of phospholipids, the main component of biological membranes, is very important from a viewpoint of structural biology and chemistry. Unilamellar vesicles are especially interesting because most biological membranes are unilamellar. On the other hand, unilamellar vesicles can be used as delivery agents; thus, the knowledge of its structure is important for pharmacology.

A standard method to investigate the form and size of vesicles is the dynamic and static light scattering. However, in this way it is impossible to obtain information about the thickness and internal structure of the membrane bilayer [1].

A more informative method is the small angle neutron scattering (SANS). The membrane bilayer thickness can be fitted from the SANS spectra on the basis of the hollow sphere model (HS) [2,3,4]. In fact, this approach suggests a uniform length density across bilayer, i.e. we have to omit the internal structure of vesicles. However, in the recent investigations [5,6], the HS approach was applied to study the internal lipid bilayer structure: the lipid bilayer was devided into hydrophilic and hydrophobic parts; each part was described by a uniform scattering length density. The HS model has two imperfections: (1) one cannot say anything about water distribution in the hydrophilic part of the bilayer; (2) it is impossible to define the place of the molecules inside the bilayer, i.e. this approach makes it impossible to study the multicomponent systems (vesicular based delivery agents of drugs for example).

The separated form-factor model (SFF) [7] looks more perspective from this viewpoint: it allows one to simulate the scattering length density by any integrable function.

In this contribution, the SFF model is used to study the structure of the polydispersed polulation of vesicles from the SANS data. We analyze the parameters of vesicles and the membrane bilayer for various hierarchic models of the length scattering density of neutrons across the membrane. We show that the water distribution in the hydrophilic part of membrane can be described by a linear function. The parameters of the vesicle population (membrane thicknesses, average radius, polydispersity, number of liner distributed water molecules in the membrane bilayer) are calculated only from the SANS spectra, without additional methods (light scattering, diffraction, etc.).

**2. EXPERIMENT**

Unilamellar dimyristoylphosphatidylcholine vesicles (DMPC) were prepared by extrusion of 15mM (1% w/w) suspension of DMPC in $D_2O$ through filters with a pore diameter of 500Å.

The SANS spectra from unilamellar vesicles at T=30°C were collected at two different spectrometers.

1. YuMO time-of-flight spectrometer of IBR-2 pulse reactor at the Joint Institute of Nuclear Research (JINR), Dubna, Russia [8]. Two sample-to-detector distances were used: 13.7m and 4.38m. The spectra were normalized on the macroscopic cross-section of vanadium [9]. Incoherent background was subtracted from the normalized cross-section of vesicles as described in [5].

2. SANS spectrometer of the Swiss Spallation Neutron Source at the Paul Scherrer Institute (PSI), Switzerland. Three sample-to-detector distances were used: 2m, 6m, and 20m. Neutron wavelength was 4.7±0.47Å. The spectra were normalized on the macroscopic cross-section of $H_2O$. The value of the incoherent background was a fitted parameter in the model calculations.

**3. THE FITTING PROBLEM IN THE FRAMEWORK OF THE SFF MODEL**

The macroscopic coherent scattering of monodispersed population of vesicles is defined by the formula [10]:

$$\frac{d\Sigma}{d\Omega}_{mon}(q) = n \cdot A^2(q) \cdot S(q) \qquad (1)$$

where $n$ is a number of vesicles per unit volume, $A(q)$ is the scattering amplitude of vesicle, $S(q)$ is vesicle structure factor ([11] describes how to include the structural factor into the model; it is also shown there that one can put $S(q) \approx 1$ for 1% concentration of DMPC); $q$ is the length of scattering vector ($q = 4\pi \sin(\theta/2)/\lambda$, $\theta$ - the scattering angle, $\lambda$ - the neutron wavelength).

The scattering amplitude in the spherically symmetric case is equal [10] to

$$A(q) = 4\pi \cdot \int \rho(r) \cdot \frac{Sin(qr)}{qr} \cdot r^2 \cdot dr. \qquad (2)$$

Here $\rho(r) = \rho_C(r) - \rho(D_2O)$ is the neutron contrast between the length scattering density of the lipid bilayer $\rho_C(r)$ and $D_2O$ ($\rho(D_2O) = 6.4 \cdot 10^{10}$ cm$^{-2}$).

Eq.(2) can by rewritten as follows:

$$A(q) = 4\pi \cdot \int_{-d/2}^{d/2} \rho(x) \cdot \frac{Sin[(R+x) \cdot q]}{(R+x) \cdot q} \cdot (R+x)^2 \cdot dx. \qquad (3)$$

Here $R$ is the radius of vesicle, $d$ is the membrane thickness.

Integration of eq.(3) in assumption $R \gg d/2$, $R+x \approx R$ gives

$$A_{sff}(q) = 4\pi \cdot \frac{R^2}{qR} \cdot Sin(qR) \cdot \int_{-d/2}^{d/2} \rho(x) \cdot Cos(qx) \cdot dx. \qquad (4)$$

Thus, the macroscopic cross-section of the monodispersed population of vesicles can be written as

$$\frac{d\Sigma}{d\Omega}_{mon}(q) = n \cdot F_s(q,R) \cdot F_b(q,d) \qquad (5)$$

where $F_S(q,R)$ is a form-factor of the infinitely thin sphere with radius $R$ [12]

$$F_s(q,R) = \left(4\pi \cdot \frac{R^2}{qR} \cdot Sin(qR)\right)^2 \qquad (6)$$

and $F_b(q,d)$ is a form-factor of the symmetric lipid bilayer.

$$F_b(q,d) = \left(\int_{-d/2}^{d/2} \rho(x) \cdot Cos(qx) \cdot dx\right)^2. \qquad (7)$$

Eqs.(5)-(7) present the separated form-factors model (SFF) for large unilamellar vesicles [7]. This model has an advantage due to the possibility to describe the membrane structure via representation of $\rho(x)$ as any integrable function. For example, for $\rho(x) \equiv \Delta\rho$=const we obtain

$$F_b(q,d) = \left(\frac{2\Delta\rho}{q} \cdot Sin\left(\frac{qd}{2}\right)\right)^2. \qquad (8)$$

The uniform density $\rho(x) \equiv \Delta\rho$=const is not realistic. Nevertheless, it is applied quite often (see, for example, [2-4]) because it allows one to restore some parameters of vesicles (average radius, polydispersity) with a good accuracy.

The vesicle polydispersity is described by nonsymmetric Schulz distribution [4,5,13]

$$G(R) = \frac{R^m}{m!} \cdot \left(\frac{m+1}{\bar{R}}\right)^{m+1} \cdot \exp\left[-\frac{(m+1) \cdot R}{\bar{R}}\right], \qquad (9)$$

where $\bar{R}$ is an average vesicle radius, $m$ is a coefficient of polydispersity. Relative standard deviation of vesicle radius is $\sigma = \sqrt{\frac{1}{(m+1)}}$.

Thus, macroscopic cross section $d\Sigma(q)/d\Omega$ has the following form:

$$\frac{d\Sigma}{d\Omega}(q) = \frac{\int_{R\min}^{R\max} \frac{d\Sigma}{d\Omega}_{mon}(q,R) \cdot G(R,\bar{R}) \cdot dR}{\int_{R\min}^{R\max} G(R,\bar{R}) \cdot dR}, \qquad (10)$$

where $R_{\min}$=100 Å, $R_{\max}$=1000 Å.

The experimentally measured cross section $I(q)$ is not completely equal to actual macroscopic cross-section $I_m(q)$ because the resolution function of the spectrometer is not a delta-function. The experimental cross-section $I(q)$ can be given by

$$I(q) = I_m(q) + \frac{1}{2} \cdot \Delta^2 \cdot \frac{d^2 I_m(q)}{dq^2}. \qquad (11)$$

where $\Delta^2$ is a second moment of the resolution function [8,14], $I_m(q)=d\Sigma(q)/d\Omega$ (see eq.(10)). Note that the resolution function for the YuMO spectrometer is well known, it is close to the Gaussian [8].

For the fitting of experimental data we used function $\chi^2$:

$$\chi^2 = \frac{\sum_{i=1}^{N}[(\Sigma_{эксп}(q_i) - I(q_i))/\delta(q_i)]^2}{N-k} \qquad (12)$$

where $\delta$ are the experimental statistical errors, $N$ is a number of experiment points, $k$ is a number of unknown parameters, $\Sigma_{эксп}$ – experimentally measured coherent macroscopic cross sections.

The fitting parameters are average vesicle radius $\overline{R}$, coefficient of polydispersity $m$, thickness of the lipid bilayer $d$, and parameters of function $\rho_C(x)$, modeling the neutron scattering length density. Beside, one can consider the incoherent background as another unknown parameter of the model. The value of the incoherent background for the case of 15mM DMPC concentration is approximately equal to 0.00546 см$^{-1}$. In the fitting the YuMO spectrometer data, this value of the incoherent background was subtracted from the experimentally measured macroscopic cross-section. For the case of PSI SANS spectrometer, the incoherent background value was a calculated parameter.

We still need to define the number of vesicles per unit volume $n$. It can be obtained by the following way. It is known that the volume of molecular DMPC in the liquid phase is equal to 1101Å$^3$ [15]. The volume of the lipid bilayer can be calculated by formula

$$V = 4\pi/3\ [(R+d/2)^3 - (R-d/2)^3]. \qquad (13)$$

So, $M=V/1101$ is the number of DMPC molecules in a single vesicle. The number of DMPC molecules in $cm^3$ can be estimated as $C=89.17\cdot10^{17}$. It means that we can put $n=C/M$. Note that $n$ is not constant, it depends on unknown fitting parameters $\bar{R}$ and $d$.

Finally, in order to estimate the fit quality, we used the following formula:

$$R_I = \frac{1}{N}\cdot\sum_{i=1}^{N}\left(\frac{\log|\frac{d\Sigma}{d\Omega}(q_i)|-\log|\frac{d\Sigma}{d\Omega}_{exp}(q_i)|}{\log|\frac{d\Sigma}{d\Omega}_{exp}(q_i)|}\right)^2. \quad (14)$$

### 4. RESULTS AND DISCUSSION.

To fit the SANS data in the framework of SFF model, the Fortran code was developed using the minimizing code DFUMIL from the JINRLIB library (JINR, Dubna).

The internal structure of the lipid bilayer was described by three types of function $\rho(x)$ given on fig.1.

In order to test the code, we calculated parameters of the unilamellar vesicles of $C_{12}E_4$. Our results [20] are in agreement with results of [5] where the same calculations were performed on the basis of the hollow sphere model.

Results of fitting the DMPC vesicles spectra of the YuMO spectrometer are given on fig.2 and Table 1. Values $\bar{R}$ and $m$ were calculated only for the uniform density and fixed for other calculations. This approach does not get worse the fit accuracy but makes a smaller number of unknown parameters of the model. Everywhere $m=10$, $\rho_{CH}=-0.36\cdot10^{10}cm^{-2}$, $S(q)=1$, incoherent background $IB=0.00546$ $cm^{-1}$.

Note that the calculated values $\chi^2$ (eq.(12)) are close to 1 ($\chi^2=1.31$ for the density (**a**) and $\chi^2=1.15$ for the cases (**b**) and (**c**)); it shows that the numerical scheme works correctly.

The introduction of the internal structure of the membrane containing hydrophobic and hydrophilic parts, leads to increasing the thickness of membrane on 5.4 Å (see variants (**a**) and

(b) of the Table 1 and fig.1). The calculated thickness of the hydrophobic part of the DMPC membrane 13.2±0.7 Å is in agreement with a result for a hydrophobic part of the POPS membrane in [5] 13±1 Å. Calculated membrane thickness 42.1±0.4 Å is a little smaller the value 44.2 Å obtained in [15] from x-ray diffraction experiment at multilamellar DMPC vesicles. Our results show that the phospholipid hydrophilic part 14.5 Å is significantly higher than the size of its polar head 9 Å [5], i.e. water molecules penetrate into the region of hydrocarbon tails on approximately 5 Å (it corresponds to the length of two methylene groups).

From our calculation, one can make a conclusion about water distribution inside hydrophilic part of the bilayer. Two variants of the linear water distribution are presented in Table 1: c1 and c2. Variant c2 corresponds to the case of density plotted on fig. 1c by a dashed line. Variant c1 (solid line on fig. 1c) shows a situation where $\rho_{PH2} = \rho_{D2O} = 6.4 \cdot 10^{10}$ cm$^{-2}$. Both variants give similar results; however, the c1 case provides a little smaller value of residual $R_I$.

Let as apply the results of variant c1 for estimation of the number of water molecules $N_W$ per one DMPC molecule penetrating into the bilayer. Assuming that all water molecules are distributed linearly across the bilayer, value $N_W$ can be calculated as follows:

$$\frac{(\rho_{PH2} - \rho_{PH1})}{2} \cdot \frac{(d - D)}{2} \cdot A = N_W \cdot l_{D2O} \qquad (15)$$

where $A$=59.6 Å$^2$ – the membrane surface square per one DMPC molecule [15], $l_{D2O}$ =1.914·10$^{-12}$ cm – scattering length of D$_2$O molecule. One can obtain from eq.(15) $N_W$ = 5.7±0.3. This value is smaller than value 8.6 obtained from the X-ray diffraction on multilamellar vesicles [15] and it is in agreement with the value 7±2 calculated from SANS in [6].

Fitting results for the spectra of the PSI SANS experiment are presented on fig.3 and Table 2. In this calculation the structure factor was included as in [11]; incoherent background

IB was not fixed beforehand but restored in the fitting. For the resolution function correcton, we used value $\Delta q/q=20\%$ at small q and $\Delta q/q=10\%$ at large q. The fit parameters for three different models of $\rho(x)$ (fig.1) are: $\overline{R}$, m, d, D, $\rho(x)$, IB.

The estimation of water quantity with linear distribution across the hydrophilic part of membrane (eq.15)) gives $N_W=3.9\pm0.03$.

It is seen that the values of the average radius given in Tables 1 and 2, are in good agreement.

As we have already mentioned, at the YuMO experiment data fitting the average radius was a fixed parameter at the vesicle structure evaluation. Contrary, in case of the PSI SANS data fitting, the value of $\overline{R}$ was a fit parameter for all the model calculations. It was possible because the experimental conditions of PSI small-angle spectrometer are better for the determination vesicle radius: at YuMO spectrometer $q_{min}=0.083$ Å$^{-1}$ while at PSI SANS spectrometer $q_{min}=0.0033$Å$^{-1}$. According to the SFF model, the possibility to measure the scattering curve in a small value of q is important for the vesicle radius evaluation (see eq.(6)). That is why the value of $\overline{R}$ can be directly fitted for any model of the scattering length density at the our calculation,

Relative standard deviation of vesicle radius $\sigma=0.30$ from experiment at YuMO spectrometer and $\sigma=0.27$ from experiment at PSI SANS spectrometer. This small difference can arise due to the accuracy of estimation of the resolution spectrometer function.

Information on the internal membrane structure obtained at two different spectrometers, shows that the calculated membrane parameters strongly depend on the used range of scattering vector. At PSI SANS spectrometer, the maximum value of q corresponds to 0.56Å$^{-1}$ (against the value 0.2Å$^{-1}$ at the YuMO spectrometer). The statistical errors at the end portion of the scattering curve of YuMO spectrometer, are large enough; in fact, this curve was measured with good statistics only to $q_{max}=0.15$Å$^{-1}$. According to the SFF model, the end of scattering curve corresponds to the form-factor of bilayer (see eq.(7)). Thus, the accuracy of restoring the DMPC membrane structure evaluation depends on the possibility to collect SANS curve in the region of large values of q, as it was done with PSI SANS spectrometer. The value of membrane thickness 47.4Å and thickness of hydrophobic membrane part 17.3Å from the PSI experiment exceed corresponding to the values 42.5Å and 11Å obtained from the YuMO experiment. The values of the hydrophilic part of DMPC membrane calculated from the PSI experiment, 15.1Å, and from YuMO experiment, 15.8Å, are in a reasonable agreement.

The last fitted parameter was the value of incoherent background (IB). The calculated values are in the range of 0.0050 – 0.0059 cm$^{-1}$ and correspond to the theoretical value for 15mM DMPC concentration, 0.0055 cm$^{-1}$.

## 5. CONCLUSION

On the basis of the SFF model, the scheme and code of fitting the SANS spectra of polydispersed vesicle population have been developed taking into account a structural factor, a spectrometer resolution function, and internal structure of vesicles.

The accuracy of the vesicle structure fitting depends on the experimentally measured range of the scattering vector. This means that the restored parameters of the internal membrane structure depend on the value of maximum q measured experimentally. For the systems under study, the best experimental conditions were realized for PSI SANS spectrometer with possibility to collect a scattering curve in the q range from $q_{min}$=0.0033Å$^{-1}$ to $q_{max}$=0.56Å$^{-1}$.

The SFF model for the SANS data of PSI spectrometer allowed one to calculate parameters of the polydispersed DMPC vesicle population: average radius 275±0.4 Å, polydispersity 27%, lipid bilayer thickness 47.4±0.04Å; thickness of its hydtrophobic and hydrophilic parts 17.3±0.05 Å and 15.05±0.09 Å, respectively. A number of water molecules per one DMPC molecule which are linearly distributed across the hydrophilic part is estimated as 3.9±0.03.

**Acknowledgements**

We are grateful to I.V.Puzynin, V.L. Aksenov, N.Popa and I.V.Amirkhanov for the useful discussion.

The investigation was supported by the Ministry of Science and Technology of the Russia Federation (grant №40.012.1.1.1148), Grant of Leading Scientific School, and RFBR (grant №03-01-00657).

Table 1. Parameters of DMPC versicles (T=30°C) calculated in the framework of SFF model for different forms of the length scattering density of neutrons across lipid bilayer (YuMO spectrometer).

| | $\bar{R}$, Å | $d, D$, Å | $\rho$, $10^{10}$см$^{-2}$ | $R_I$, % |
|---|---|---|---|---|
| (a) | 277±5 | 36.7±0.1 | $\Delta\rho$=5.1±0.01 | 1.3 |
| (b) | 277(фикс.) | 42.1±0.4 <br> 13.2±0.7 | $\rho_{PH}$=2.5±0.1 | 2.6 |
| (c1) | 277(фикс.) | 42.5±0.3 <br> 11.0±0.9 | $\rho_{PH1}$=4.1±0.1 <br> $\rho_{PH2}$=6.4(фикс) | 2.4 |
| (c2) | 277(фикс.) | 42.7±0.4 <br> 12.7±0.9 | $\rho_{PH1}$=4.4±0.09 <br> $\rho_{PH2}$=5.4(фикс.) | 2.5 |

Table 2. Parameters of DMPC vesicles (T=30 °C) calculated in the framework of SFF model for different forms of the scattering length density of neutrons across lipid bilayer (PSI SANS spectrometer).

| | $\bar{R}$, Å | $m$ | $d, D$, Å | $\rho$, $10^{10}$cm$^{-2}$ | IB, $10^{-3}$cm$^{-1}$ | $R_I$, % |
|---|---|---|---|---|---|---|
| (a) | 272.9±0.4 | 12 | 36.7±0.021 | $\Delta\rho$=4.91±0.005 | 5.01±0.01 | 0.55 |
| (b) | 275.3±0.4 | 13 | 46.4±0.03 <br> 18.1±0.03 | $\rho_{PH}$=3.4±0.003 | 5.76±0.01 | 0.16 |
| (c) | 275.0±0.4 | 13 | 47.4±0.04 <br> 17.3±0.05 | $\rho_{PH1}$=4.9±0.001 <br> $\rho_{PH2}$=6.4(fixed) | 5.899±0.01 | 0.15 |

**FIGURE CAPTURES**

Fig.1 (a) – the uniform length scattering density; (b) – the 'step' length scattering density; (c) – the density of the linear function type. $\rho_{D2O}$, $\rho_{PH}$, $\rho_{CH}$ – the scattering length density of the $D_2O$, hydrophilic and hydrophobic parts of lipid bilayer, respectively. D and d are, respectively, thickness of hydrophobic part of membrane and thickness of lipid bilayer.

Fig.2. Results of fitting of DMPC vesicle spectrum for three variants of the internal structure of lipid bilayer given at fig.1 (YuMO spectrometer)

Fig.3 Results of fitting of DMPC vesicle spectrum for three variants of the internal structure of lipid bilayer given at fig.1 (PSI SANS spectrometer)

Fig.1

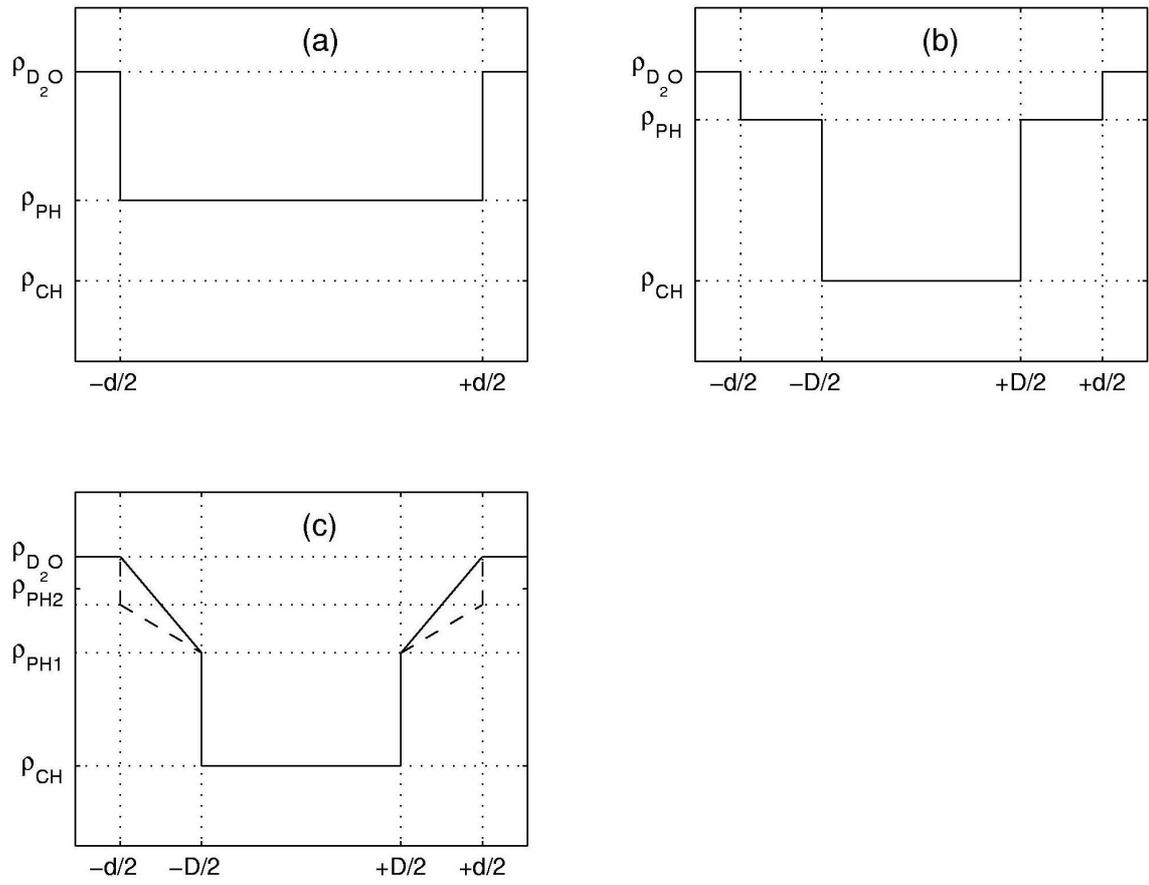

Fig.2

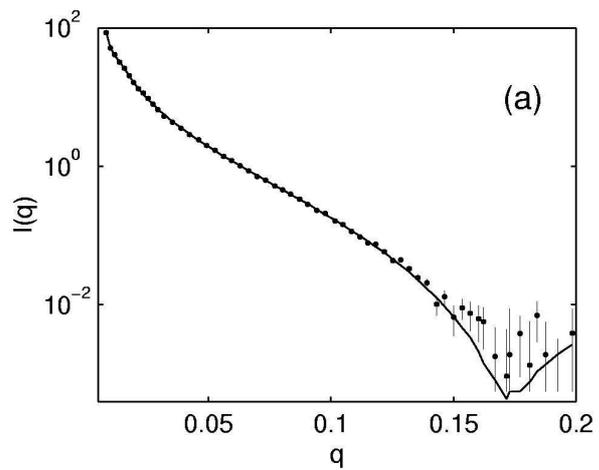
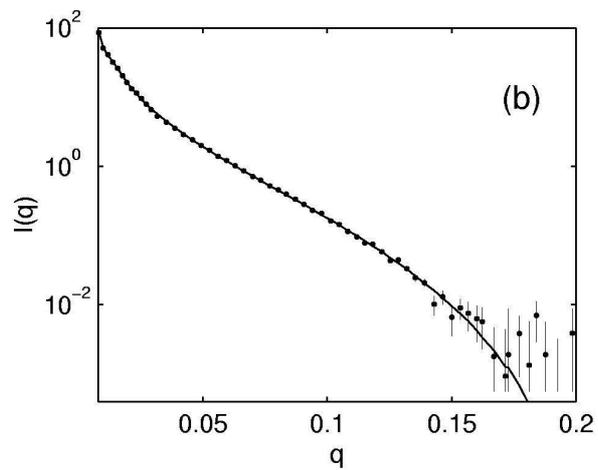
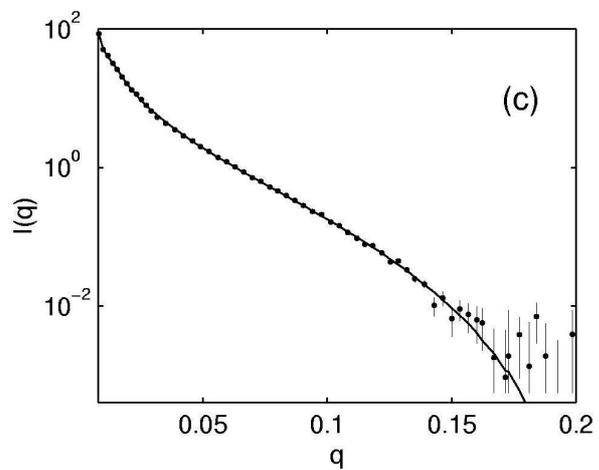

Fig.3

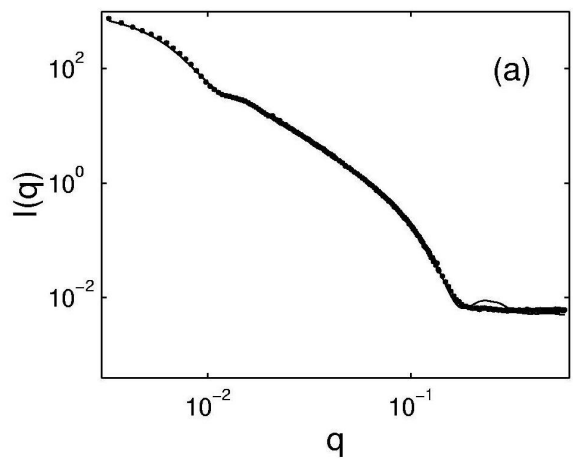

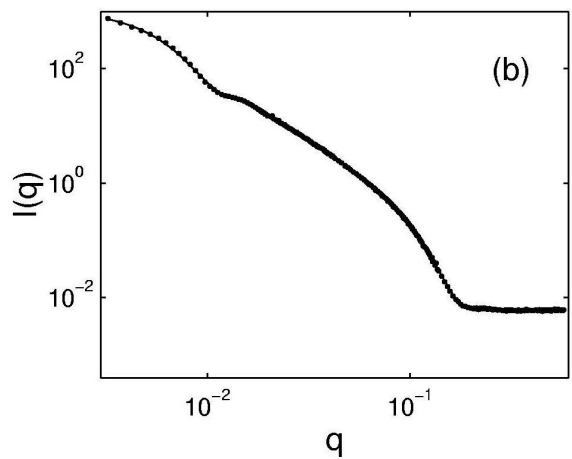

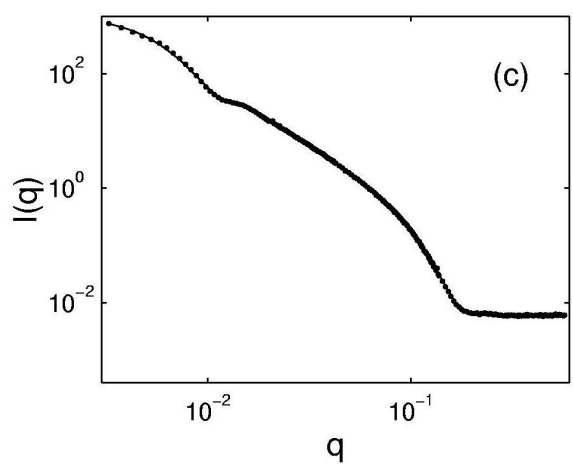